\documentclass{svjour3}                     
\smartqed  
\usepackage{graphicx}
 \usepackage[cmex10]{amsmath}
\begin{document}

\title{Community Structure and Patterns of Scientific Collaboration in Business and Management
}

\titlerunning{Communities and Patterns of Scientific collaboration}        

\author{T.S. Evans         \and R. Lambiotte \and P. Panzarasa 
}


\institute{T.S. Evans \at
              Imperial College London, Physics Department and Complexity \& Networks\\
           \and
           R. Lambiotte \at
              Imperial College London, Institute for Mathematical Sciences\\
            \and
            P. Panzarasa \at
              Queen Mary University of London, School of Business and Management \\
}

\date{Accepted 16 May 2011 for publication in Scientometrics.}

\maketitle

\begin{abstract}

\textbf{This paper investigates the role of homophily and focus constraint in shaping collaborative scientific research. First, homophily structures collaboration when scientists adhere to a norm of exclusivity in selecting similar partners at a higher rate than dissimilar ones. Two dimensions on which similarity between scientists can be assessed are their research specialties and status positions. Second, focus constraint shapes collaboration when connections among scientists depend on opportunities for social contact. Constraint comes in two forms, depending on whether it originates in institutional or geographic space. Institutional constraint refers to the tendency of scientists to select collaborators within rather than across institutional boundaries. Geographic constraint is the principle that, when collaborations span different institutions, they are more likely to involve scientists that are geographically co-located than dispersed. To study homophily and focus constraint, the paper will argue in favour of an idea of collaboration that moves beyond formal co-authorship to include also other forms of informal intellectual exchange that do not translate into the publication of joint work. A community-detection algorithm for formalising this perspective will be proposed and applied to the co-authorship network of the scientists that submitted to the 2001 Research Assessment Exercise in Business and Management in the UK. While results only partially support research-based homophily, they indicate that scientists use status positions for discriminating between potential partners by selecting collaborators from institutions with a rating similar to their own. Strong support is provided in favour of institutional and geographic constraints. Scientists tend to forge intra-institutional collaborations; yet, when they seek collaborators outside their own institutions, they tend to select those who are in geographic proximity. The implications of this analysis for tie creation in joint scientific endeavours are discussed.}
\keywords{Collaboration networks \and Community structure \and Intra- and inter-institutional collaborations \and Geographic distance \and Research specialty}
\end{abstract}

\section{Introduction}
\label{intro} The idea of using published papers to study collaboration patterns among scientists is not new \cite{price65}. In information science, for example, there is a substantial body of literature concerned with co-authorship networks \cite{ding99} and co-citation networks \cite{crane72}, where connections between authors are defined, respectively, in terms of collaboration on the same paper or citation of their work in the same literature. Studies of scientific collaborations have even a longer history in the mathematics community, in which one of the earliest attempts to map and investigate the structure of social interaction within a scientific community was formalised through the concept of the Erd\"{o}s number, a measure of a mathematician's distance, in bibliographical terms, from the Hungarian scholar \cite{decastro99}. Only recently, however, due to the advent of new technological resources and the availability of comprehensive online bibliographies, have a number of much larger and relatively complete and detailed collaboration networks been documented and analysed \cite{barabasi02,jones08,moody04,newman01,wuchty07}.

While most of these recent studies have been interested either in the global structural and dynamic properties of the collaboration networks \cite{barabasi02,moody04,newman01}, or in the effects that collaboration has on scientific performance \cite{jones08,wuchty07}, only little attention has been given to the micro mechanisms underpinning the way scientists select their collaborators at the local level. For example, while it has been documented that collaborations spanning multiple universities, and in particular, among these, the collaborations involving solely elite universities are more likely to result in more highly cited papers than other forms of collaborations \cite{jones08}, it still remains to be explored how in reality scientists assess potential partners and select them for collaborative relations. While consideration of performance will certainly have some impact on the way collaborations are forged, it is also true that only a minority of scientists may be in a position to freely collaborate only with those that can help them achieve the highest levels of performance. For the majority of scientists, there may be structural, disciplinary, institutional, or geographic constraints that restrict their search behaviour to a delimited subset of possible collaborators. Focusing on the principles that are conducive to the highest levels of scientific performance, therefore, does not help understand how ties are actually forged in a collaboration network. To this end, what is needed is an approach to tie creation that uncovers the mechanisms that underlie the selection of scientific collaborators, irrespective of their implications for performance.

In this paper, we take a step in this direction, and uncover the role of two fundamental mechanisms of tie creation in collaboration networks: homophily \cite{lazarsfeld54,mcpherson01} and focus constraint \cite{feld81}. We examine whether scientists adhere to a principle of exclusivity in selecting their collaborators, by choosing only among those with whom they share similar attributes. We focus on two forms of homophily. First, scientists may take the research specialty of potential intellectual partners as cues, and select those with whom there is a substantial overlap of research interests, scientific background, practices, perspectives, and standards. Second, when there is uncertainty on the scientific quality of a joint work, scientists may choose to affiliate themselves with others whose status is similar to their own \cite{podolny94}.

We then shift our attention to focus constraint, and examine the extent to which institutional and geographic constraints govern the creation of collaborative ties. First, scientists may be more likely to select collaborators with whom they share the same institutional affiliation than others from different institutions. Second, intra-institutional collaborations may be induced by the tendency of scientists to collaborate with others that are geographically co-located. This tendency would also imply that, when collaborations span different institutions, they are more likely to involve scientists that are in geographic proximity than at long distances from one another.

In our study we also attempt to adopt a broader perspective on collaboration than the one strictly implied by the idea of co-authorship. Co-authorship undoubtedly represents one of the major forms of intellectual cooperation. The literature, however, has long argued in favour of a more permeable concept of collaboration to include other forms of informal interaction among scientists \cite{katz97,laband00}. For instance, the published work of an author typically benefits from comments provided by colleagues, journal reviewers and editors. Other forms of informal intellectual collaboration include the mentoring that senior scientists offer to junior ones, and the commentary received during the presentation of papers at conferences, workshops, and professional meetings. Moreover, it is not uncommon that scientists become indirectly connected as a result of collaborative agreements between higher-level units, such as departments, institutions, and research centres \cite{katz97}. For instance, team leaders might agree on a common research agenda that commits their respective groups to a number of collaborative endeavours. In this case, while certain members of the groups may not be directly involved in joint work leading to formal co-authorship, nonetheless their research may indirectly benefit from the transfer of knowledge and skills, cross-fertilisation of ideas, and the establishment of common research standards and goals that the collaborative agreement between their groups has made possible. In cases like these it is not always an obvious task to identify who is collaborating with whom precisely because patterns of co-authorship and collaboration tend to diverge. While a strict bibliometric assessment would count as collaboration only those activities that translate into a joint paper, there are certainly other peripheral or indirect forms of intellectual exchange that are not reflected in formal co-authorship, and yet represent genuine instances of associations that should be taken into account to adequately capture the full extent of scientific collaboration.

To undertake an accurate assessment of collaboration one would therefore need to integrate data on formal co-authorship with details on informal commentary \cite{laband00}. This would inevitably be an arduous task, especially when conducted on a large scale. Here, we propose an alternative response to the problem of the opaqueness of collaboration. We begin by constructing a collaboration network based on formal co-authorship, in which, as is typically done in similar network studies, two scientists are assumed to be connected when they appear among the authors of the same paper \cite{barabasi02,newman01}. However, we move beyond the idea of dyadic direct connections between scientists, and apply recent community-detection methods to partition the network into communities \cite{fort}. A scientist belongs to a community when he or she collaborates with other members of that community to a greater extent than with members of other communities. In this sense, communities may be locally dense even when the network as a whole is sparse. Moreover, because within each community scientists are inevitably connected only to a subset of all other members, communities may include scientists that are only indirectly connected with each other.

We study the role of homophily and focus constraint for \textit{all} scientists within each community, even those that are not directly connected with each other. In so doing, we implicitly take on a two-fold perspective on the structure and meaning of collaboration. First, we assume collaboration occurs only within but not across the boundaries of communities. Second, while direct ties clearly reflect formal co-authorship, we regard indirect ties as an indicator of informal forms of collaboration. We seek support in favour of this perspective by examining the collaboration network based on the papers submitted to the 2001 Research Assessment Exercise (RAE) in the UK in the field of Business and Management. Drawing on accurate details on the scientists' attributes, we examine the extent to which the topological boundaries between the uncovered communities reflect some fundamental ways in which scientists collaborate, either formally or informally. We do this by testing the tendency of communities to include pairs of scientists that work in the same research specialties, are affiliated with the same institutions, are associated with the same levels of status, and are located at geographic proximity with each other.

The rest of the paper is organised as follows. In the next section, we place our work within the relevant theoretical context. We then introduce the data and the methods for partitioning the network into communities and assessing homophily and constraint in each community. In Section \ref{sec:results}, we present the results. The final section will summarise and discuss the main findings.

\section{Homophily and focus constraint in collaboration networks}
\label{tie}

Homophily represents one of the network mechanisms of tie creation with the longest tradition of investigation in the social sciences. This is the principle that similarity breeds connection \cite{lazarsfeld54,mcpherson01}. A significant body of research has provided supportive evidence in favour of homophily by documenting a positive association between sharing an attribute and some baseline level of interpersonal attraction \cite{mcpherson01}. Attraction could, in turn, be reflected in a heightened probability of similar people to select each other \cite{kossinets06}, or communicate more frequently and develop a stronger social interaction \cite{reagans05}.

In this paper, we begin our investigation of homophilous interactions in collaboration networks by examining the extent to which scientists that work in the same research specialty collaborate with one another with a higher likelihood than scientists from different specialties. While research has long been interested in assessing the benefits of conducting research across disciplinary fields and research specialties \cite{laband00,whitfield08}, the fact that scientists can also develop dense and strong connections within their own fields or specialties has often received scanty attention. Scientists can carefully select their collaborators to draw on different knowledge pools without having to acquire the needed knowledge personally, but they can also aim to strengthen their skills and enhance scientific consensus within their own specialty area. Recent work suggests that scientists embedded in collaboration networks share ideas, scientific standards and technique \cite{moody04,whitfield08}. By selecting their collaborators within their own specialty area, scientists can enhance scientific cohesion and embeddedness, receive validation of their own attitudes and beliefs, and facilitate their scientific production through the generation of shared norms of research practice.

A second manifestation of homophilous interactions in collaborative research is related to the role of status similarity in tie creation. In the social sciences, a number of empirical studies have long been interested in the processes and reasons underpinning the creation of connections among economic actors of similar status. Research has shown that processes of competitive isomorphism are likely to lead economic actors of similar status to adopt similar practices and operating systems, which in turn facilitates the coordination of cooperative activities \cite{chung00,lorange92}. Status similarity also aligns the expectations of potential partners about each other's behaviour, and increases their commitment to sharing both the costs and benefits of an interaction \cite{chung00}. Moreover, a substantial body of work in sociology has illustrated that economic actors, when considering the choice of creating a connection, assess the status of potential partners \cite{chung00,podolny94}. For example, the way in which others perceive the quality of the output of a firm, especially when it cannot be assessed without ambiguity, depends on the status of other firms that interact with the focal firm \cite{podolny94}. As a result of the signaling effect of status positions and social interactions, firms with similar status tend to establish connections with one another when there is uncertainty about the output of their transaction.

Sociological research on culture, science and technology has proposed a similar view on the relational foundations and signaling effect of status. For instance, it was found that within artistic genres with limited objective standards and high levels of uncertainty on quality, the perception and judgement of the work of an artist was contingent on the status of other artists with whom the focal artist interacted in the artistic community \cite{greenfeld89}. In the sociology of science, it was contended that, when there are pronounced levels of uncertainty about scientific quality, such as during periods of paradigmatic change, the way in which a scientist is regarded depends on the status of those with whom the scientist is associated \cite{camic92,latour87}. A similar perspective was also suggested to explain the development of technology, in the sense that when an inventor's technology cannot be evaluated without uncertainty, assessment is fundamentally based on the status of the economic actors that endorse that technology \cite{podolny95}.

These studies thus suggest that a principle of exclusivity based on status may also govern the selection of partners in scientific collaborations. Challenged by pronounced levels of competitive pressure and uncertainties posed by the need to secure funding and publish in high-quality journals, scientists will become increasingly exclusive in the formation of collaborations. They will generally avoid collaborating with others of a lower status, and instead select collaborators of roughly equivalent status \cite{jones08}. In this paper, to investigate status-based homophily, we measure the ranking of the institutions with which scientists are affiliated, and then examine whether collaborations tend to span institutions of different ranking or only those with a similar one.

The second ordering principle that we examine is focus constraint \cite{feld81}. This refers to the idea that social associations depend on opportunities for social contact. Research has uncovered the tendency of connections to occur among individuals who share activities, roles, social positions, institutional affiliations, and geographic location \cite{feld81,kossinets06,monge85}. Here a special emphasis is placed on institutional and geographic constraint. First, we examine whether scientists are more likely to establish collaborations within their own institutions than across institutional boundaries. Recent studies have investigated forms of collaborations that involve organisations of various institutional profile, such as academic departments, business firms, government and non-government organisations \cite{leydesdorff05}. In particular, research has highlighted the role of these inter-institutional collaborations in sustaining knowledge transfer and creation. There are, however, also benefits associated with intra-institutional collaborations. In principle, scientists can choose their collaborators within their own institutions for a variety of reasons. For instance, joint research may be facilitated by the ease and frequency of face-to-face communication and meetings, and by the common cultural orientations, scientific standards and practices that are typically shared by the members of the same institution. Hiring policies, in turn, can also promote collaborative research within institutions as they tend to emphasise overlapping areas of research interests between applicants and incumbents leading to potential joint work. For these reasons, here we examine the role of institutional constraint in scientific collaboration by testing the tendency of scientists to restrict the choice of their partners within institutional boundaries.

Intra-institutional collaborations may also originate from the benefits that scientists gain from being geographically close to one another. The literature has long investigated the benefits of geographic proximity, and in particular its impact on innovative activities \cite{jaffe93}. Even though knowledge could in principle travel through space inexpensively, nonetheless knowledge production tends to be geographically clustered \cite{braunerhjelm}. The arguments often proposed to explain this phenomenon include the benefits that geographic proximity offers in terms of knowledge spillovers \cite{jaffe93}, opportunities of face-to-face interaction, transfer of tacit knowledge, and the occurrence of unanticipated encounters between individuals \cite{gertler03}. While the literature has been concerned mainly with the spatial distribution of economic activities, it may also help gain a better understanding of the geography of scientific collaboration \cite{jones08}. When selecting their collaborators, scientists may be encouraged to choose them within short geographic distances because spatial proximity facilitates informal communication and the transfer of complex knowledge, which in turn may lead to an increasing commitment to cooperation \cite{katz97}. This argument thus not only suggests that scientific collaboration may tend to occur within institutional boundaries, but also that, when collaborations span different institutions, they may be more likely to involve scientists from institutions that are geographically close than dispersed.

\section{Data and methods}

In this section, we will begin by introducing the RAE network dataset, and then the measures for scientists' attributes that will be used to study homophily and focus constraint. We will then present the community-detection algorithm that will be used to partition the network into groups of indirectly connected scientists. The section will end with a discussion of the statistical methods developed to assess homophily and focus constraint in each community.

\subsection{The data}
\label{data}

For our analysis, we have constructed the collaboration network of the social scientists that authored or coauthored the publications submitted to the RAE 2001 in Business and Management in the UK. The RAE was established in the UK in 1986, when the government introduced the policy of selective funding \cite{ball04,cooper98,hero}. The exercise is traditionally carried out by the UK government through Higher Education Funding Councils, and represents a peer-review evaluation process undertaken by panels consisting of members who are chosen by the funding bodies according to their research experience. The RAE that took place in 2001 represents the broader context from which our data were drawn. On the whole, it consisted of $68$ units of assessment and around $213,000$ publications examined. In this work, we restrict our analysis to the unit of assessment that received the largest number of submissions. This was the Business and Management Studies subject area, which received $97$ submissions from $94$ institutions \cite{ball04,hero}. Each institution was invited to put forward within its submission all individuals who were actively engaged in research and in post on 31 March 2001. Each of these individuals was required to submit up to four pieces of research output produced during the period 1 January 1996 to 31 December 2000.

Panels composed of expert academics were formed to assess the quality of submissions \cite{baker02,cooper98}. Evaluation criteria for each unit of assessment were published by the panels before submissions were made to ensure that academics were informed of the aspects of submissions that the panels regarded as most important as well as the areas on which institutions were required to comment in their submissions \cite{hero}. Ratings were allocated to submissions, and ultimately to universities, on the grounds of their ability to reach national or international levels of excellence. Ratings of research quality were expressed in terms of a standard scale including 7 points ranging from 1 to 5* (i.e., 1, 2, 3b, 3a, 4, 5, and 5*). The RAE aimed to ensure that institutions that produced research of the highest quality were allocated a higher proportion of the available funding than institutions with lower-quality research. In the RAE 2001, for example, institutions that acquired a rating of 1 or 2 did not obtain any funding, while institutions that received a rating of 5* were given four times as much funding as the institutions with a rating of 3b. The allocation of funding according to research quality was therefore intended to act as an incentive both to protect and develop research of excellent quality in the UK.

Our data contain detailed information about each paper that was submitted to the RAE 2001 in Business and Management, including the paper title, the names of author and co-authors, the RAE ratings of their institutions as well as the publication type and publishing details. Among the advantages of this dataset over other sources of data on publications is that disambiguation of institutional affiliations of the authors who submitted to the RAE is relatively straightforward. Our sample includes $9,325$ papers submitted to the RAE by $2,609$ scientists. These papers were also co-authored by $5,752$ scientists that did not submit to the RAE. Thus, the total number of scientists in our sample amounts to $8,361$. A tie is established between two scientists if they have co-authored one or more papers. Following \cite{newman01a}, the weight of a tie between two scientists reflects their contributions in their collaboration: the larger the number of scientists collaborating on a paper, the weaker their interactions. Thus, tie weight increases with the total number of papers co-authored, and is inversely proportional to the total number of co-authors of those papers. In our analysis we looked at the largest connected component of this weighted network which contains $3,338$ authors.\footnote{The next largest component has fewer than $100$ authors.}

\subsection{Scientists' attributes}
\label{NA}

To study the role of homophily and focus constraint, we needed a number of additional attributes for the scientists. Because these attributes were available only for the scientists who submitted to the RAE (and not, for example for non-UK scientists or UK PhD students who co-authored with someone who submitted, but did not submit themselves), we then had to extract the subset of these scientists from the largest connected component of the network. Of the $3,338$ scientists in the component, only $973$ submitted to the RAE. For each of these $973$ scientists, we measured research specialty, status, institutional affiliation, and geographic location.

To assess research-based homophily, we assigned each scientist to a research specialty by using the domain statements of the $24$ divisions and interest groups identified by the Academy of Management. For each of these divisions and groups, the Academy provides a brief description of the main research topics, objectives and methods.\footnote{Descriptions of these divisions and groups are available at the website of the Academy of Management: \texttt{http://www.aomonline.org/aom.asp}} By using an algorithm, we matched the titles of the papers submitted to the RAE with the Academy's statements, and assigned each author to a unique research specialty \cite{whitfield08}.

To assess status-based homophily, each scientist was assigned the RAE ranking acquired by the institution with which he or she was affiliated. Two measures of status were obtained by using the RAE ratings that institutions received in 1996 and 2001. To study geographic constraint, we obtained the latitude and longitude values in degrees for each institution, and then calculated the distance in kilometers between any pair of institutions. The geographic distance between any two scientists was then assumed to be equal to the distance between the two institutions with which the scientists were affiliated. Finally, for institutional constraint, scientists were associated with their respective institutions of affiliation.

\subsection{Community detection}

The detection of communities, or modules, in networks has attracted much attention in the last few years. Modules are defined as sub-networks that are locally dense even though the network as a whole is sparse. They have been observed in a variety of networks (e.g., biological networks, brain functional networks, and collaboration networks), where they usually correspond to functional sub-units, namely sets of nodes that have a (usually unknown) property or function in common. This architecture is  expected to naturally emerge in groups of interacting scientists \cite{andrea2}, as it presents the advantage of combining two types of social organisation \cite{inform}: {\it close} networks which foster trust and facilitate the transfer of complex and tacit knowledge, and {\it open} networks which are rich in structural holes and facilitate knowledge creation and information diffusion. Several methods have been developed to detect modules in large networks, and they cover a broad range of concepts and implementations  \cite{fort}. In the field of Scientometrics, a division of citation or collaboration networks into communities has been used as a taxonomic scheme in order to map knowledge domains \cite{borner2003,boyack,chen03,loet,rosvall,wallace}, but also as way to track their temporal changes and the mobility of researchers \cite{iina}.

In this paper, we adopt a partitioning-based viewpoint, as we look for non-overlapping communities. Partitions are uncovered by optimising the multi-resolution modularity introduced by Reichardt and Bornholdt \cite{reichardt0}:
\begin{equation}
\label{reichardt} Q(\gamma) = \frac{1}{2m} \sum_{C \in \mathcal{P}} \sum_{i,j \in C} \biggl[ A_{ij} - \gamma \frac{k_i k_j}{2m} \biggr],
\end{equation}
where $A$ is the weighted adjacency matrix of the collaboration network, $k_i \equiv \sum_{j} A_{ij}$ is the strength of node $i$ and $m \equiv \sum_{i,j} A_{ij}/2$ is the total weight in the network. The summation over pairs of nodes $i,j \in C$ belonging to the same community $C$ of the partition\footnote{Here $\mathcal P$ is a partition of the vertices of our graph. That is, $\mathcal P$ is a set of communities $\mathcal{C}$ and every author in the largest connected component of our full weighted co-authorship graph is in one but only one of these communities.} $\mathcal P$ counts intra-community links. This quality function measures if links are more abundant within communities than would be expected on the basis of chance, and incorporates a resolution parameter $\gamma$ allowing to tune the characteristic size of the modules. $Q(1)$ corresponds to Newman-Girvan modularity \cite{NG}. The resolution parameter $\gamma$ is essential in order to get rid of the size dependence of modularity and to uncover the true multi-scale organisation of the network. In what follows, the optimisation of $Q(\gamma)$ is performed by using a reliable greedy algorithm \cite{Blondel}. \footnote{The java code used to perform the optimisation of $Q(\gamma)$ is available on request from Tim Evans.}

\subsection{Statistical significance of module attributes}
\label{diversity}

By definition, uncovered modules consist of groups of scientists that are indirectly connected but are close in a topological sense. Modules thus provide coarse-grained levels of interactions which allow us to go beyond known dyadic connections between scientists present in the data and to uncover intermediate units (building blocks) from the organisation of the collaboration network. It is also important to emphasise that scientists are expected to be driven by antagonistic forces, e.g. geographic distance vs research specialty, in their choice of collaboration. The non-overlapping organisation imposed by the partitioning algorithm is thus expected to highlight the dominant factors, namely it uncovers communities underpinned by one dominant mechanism.

In order to test the effect of homophily and focus constraint on scientific collaborations, we look at two measures of attribute diversity within each community:
\begin{equation}
\label{entropy} S_C = - \sum_{v \in \Gamma} p_{c;v} \ln(p_{c;v}) ~~{\rm and} ~~ R_C = 1- \sum_{v \in \Gamma} p_{c;v}^2,
\end{equation}
where $p_{c;v}$ is defined as the density of authors in community $C$ who possess attribute $v$ in the set $\Gamma$ of possible attributes. $S_C$ and $R_C$ are the Shannon entropy and the Simpson diversity index of $p_{c;v}$, respectively. By construction, $S_C$ and $R_C$ are measures of the diversity of a certain set $\Gamma$ of attributes within community $C$. Low values of $S_C$ and $R_C$ correspond to communities whose nodes are affiliated with the same institution, work in the same specialties or are associated with the same levels of status, respectively.

Different sets of attributes are considered in order to assess the salience of different factors for community structure: institution, research specialty and RAE rating. For research specialty, for instance, (\ref{entropy}) becomes $S_C =- \sum_{v=1}^{24} p_{c;v} \ln(p_{c;v})$, where $p_{c;v}$ is now the density of authors with research specialty $v$ in community $C$ and the summation is performed over the set of $24$ possible research specialties.  The significance of these diversity measures is evaluated through a permutation test \cite{mason2}, namely by measuring $S_{C;\alpha}$ and $R_{C;\alpha}$ for each community $C$ on $1,000$ different instances $\alpha$ where the assignment of the nodes to communities is preserved but where the attributes of the nodes are randomly re-shuffled. The diversity of community $C$ is then assessed by comparing $S_C$ ($R_C$) to the value of diversity of the null models and by measuring the probability $P_{c}$ that community $C$ is less diverse than the one observed in the null model (see Fig.~1).

The salience of geographic proximity for community structure is assessed as follows. For each community, we look at two average distances: the average distance $d_\mathrm{UIP}$ between an author and all other authors in the same community provided they are not from the same institution, and the average distance $d_\mathrm{UAP}$ between an author and all authors in the same community whatever their institution. There is almost no difference in the results obtained from these two distance measures in terms of the comparison of the null models to the actual average distance measured in communities. The important point is that these distances are measured regardless of whether or not scientists co-authored a paper. Moreover, a separation of 100km when one institution is in a relatively sparsely populated location with few institutions (e.g., Northern Island) may be a short scale whereas 100km may be a comparatively large distance in a dense urban environment with many institutions. Therefore, these distances have to be compared to an appropriate null model defined as follows. Each author in a community is considered in turn. The locations of all the institutions except for the one associated with the author being considered are shuffled. Authors in the same institution thus remain in the same institution, but the distance from the author under consideration to those in another institution will almost certainly change. We calculate the average distance between all pairs of authors in the same community in $1,000$ realisations of the null model and compare the range of average distances found in the null model against the average distance measured for the community with institutions in the real location.

\section{Results}
\label{sec:results}

\begin{figure}
  \includegraphics[width=0.77\textwidth]{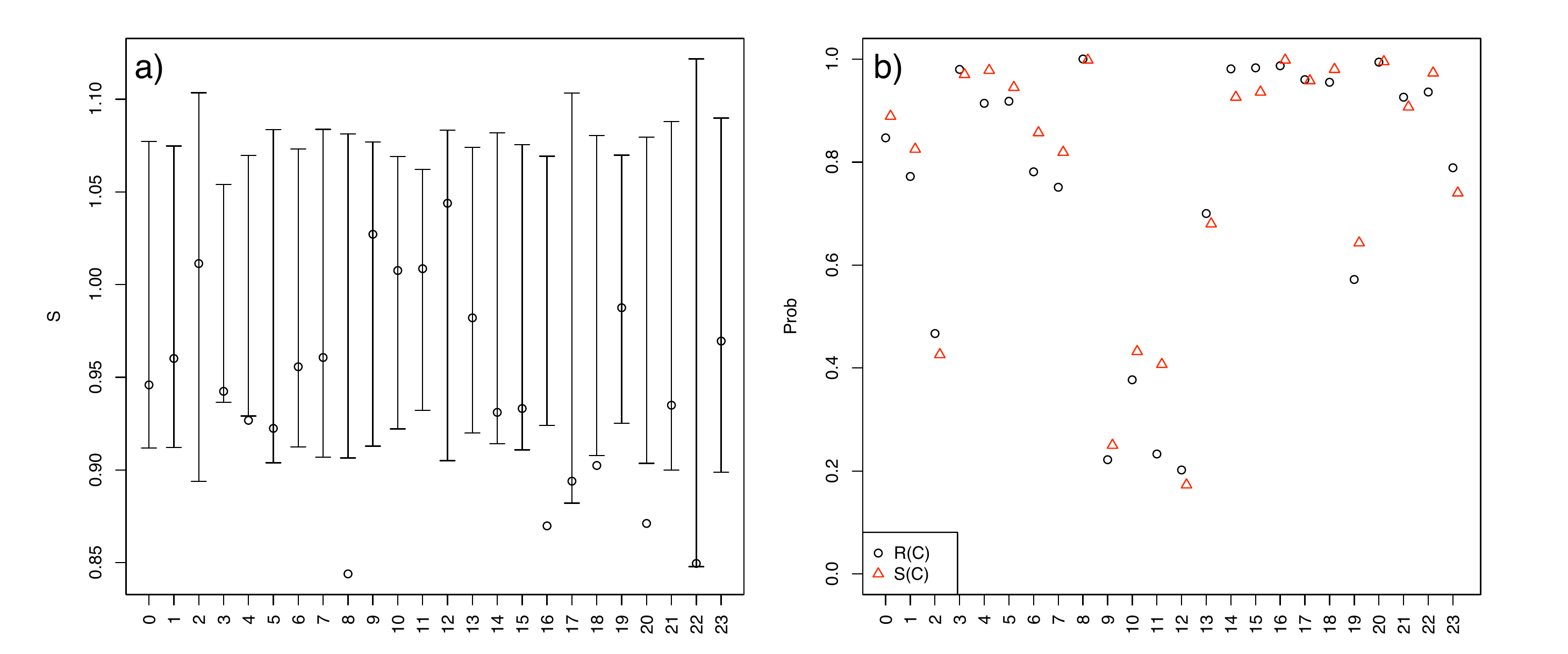}
\caption{Statistical significance of the diversity of research specialties in Business and Management for a partition of $24$ communities. (a) Points show the entropy $S_C$ of the modules for the research specialty variable. Comparison is against $1,000$ different instances of the null model described in the text. The ends of the bars mark the entropies at the quantiles $2.5\%$ and $97.5\%$. (b) Points show the probability $P_C$ that the diversity found in a null model is greater than the one found in reality.}
\label{fig:3}       
\end{figure}

Our analysis was performed on the largest connected component of the weighted collaboration network defined in \ref{data}. Modules at different scales have been uncovered by optimising $Q(\gamma)$ over a broad range of values of $\gamma$. In what follows, we will discuss the properties of the partition optimising $Q(0.091)$, keeping in mind that similar conclusions are also obtained for other values of $\gamma$. Results are similar for both diversity measures discussed in \ref{diversity}, and therefore we will only show entropy in our figures. The obtained partition is made of $24$ modules, and has been chosen to coincide with the number of research specialties. Our main purpose is to compare this algorithmically-obtained partition to our information about the scientists, namely their research specialty, RAE rating, institutional affiliation, and geographic distance.

To investigate the mechanisms driving the formation of communities, we measured the diversities $S_C$ and $R_C$ for the first $3$ sets of attributes. Community $C$ is said to exhibit a significant uniformity (lack of diversity) for a certain set of attributes if it is less diverse than in $97.5\%$ of the null models, i.e., $P_C>0.975$ in the above notations. In that case, the attributes of $C$ are thus significantly different from a random assignment. On the contrary, the composition of a community is not distinguishable from a random assignment for values of $P_C<0.975$. Geographic distance is said to be a significant factor underpinning the composition of community $C$ if the average distances $d_{\rm UIP}$ and $d_{\rm UAP}$ between its scientists are smaller than in the null model in $97.5\%$ of the random realisations.

The analysis incorporates four sets of results. The first two test our hypothesis of specialty- and status-based homophily, respectively. As shown in Fig.~\ref{fig:3}, research specialty is weakly correlated with community structure. Only $5$ communities out of $24$ exhibit a degree of homogeneity in research specialty that is statistically significant. The rest of the communities are not statistically significantly different from what would be randomly expected. These findings thus provide only partial support in favour of the hypothesis that in Business and Management scientists tend to collaborate with others within their own research specialty. At the same time, results also suggest that scientists do not work across research specialties to a greater degree than by chance. For instance, while Fig.~\ref{fig:3}b indicates that a few communities have a large probability (close to $1$) of exhibiting a greater research similarity than the one found in the null model, there is no community for which the probability that the corresponding null model has a higher research diversity is close to zero.

\begin{figure}
\includegraphics[width=0.77\textwidth]{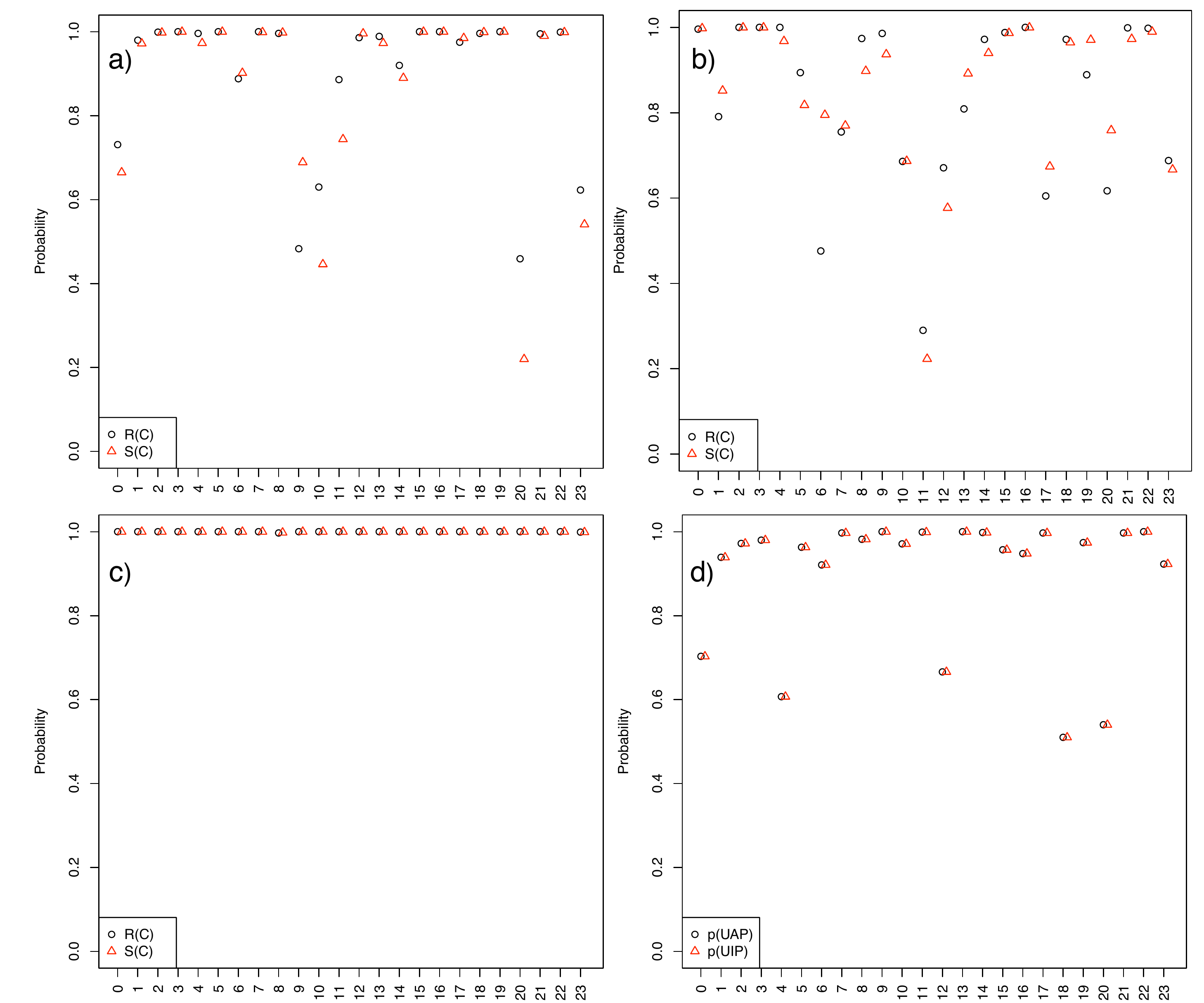}
\caption{Statistical significance of the diversity of 1996 and 2001 RAE assignments (a and b), institutions (c) and geographic distance (d) for a partition of $24$ communities. For each community, we plot the probability that diversity (a,b and c) or average distance (d) found in a null model is greater than the one found in reality.
}
\label{fig:4}       
\end{figure}

The second set of results is concerned with status homophily, namely the hypothesis that scientists tend to collaborate with others that are affiliated with institutions with the same RAE rating as their own. As shown by Fig.~\ref{fig:4}a,b, the salience of status homophily for collaboration depends on which measure of status is used. While the 1996 RAE rating appears to be a statistically significantly strong driver of collaboration for $13$ communities, similarity in the 2001 rating is correlated with collaboration only for $6$ communities. This should not be surprising. On the one hand, when scientists selected their collaborators, they were aware of the RAE rating that institutions obtained in 1996. In this respect, the results provide support to the hypothesis that scientists in most communities used the 1996 RAE rating as a signal to infer the quality of potential collaborators and discriminate between them. On the other, since the papers in our dataset were published before 2001, the RAE ratings obtained in 2001 were obviously not available to the scientists at the time of their collaboration. Thus, the 2001 RAE ratings could not have been used before 2001 to make inferences about quality, which explains the weaker support that Fig.~\ref{fig:4}a,b provides to homophily based on the 2001 rating than on the 1996 one. Due to the (weak) correlation between the 1994 and 2001 ratings, some of the scientists that before 2001 chose collaborators with a status similar to their own continued to maintain such similarity when the new RAE ratings were released in 2001. However, Fig.~\ref{fig:4}a,b suggests that there were also a number of scientists who changed their status in 2001, and as a result some of the similarities based on the 1996 ratings eventually disappeared in 2001.

The last two sets of results test the hypotheses of institutional and geographic constraints, respectively. As can be seen in Fig.~\ref{fig:4}c, communities are extremely uniform in terms of the institutional affiliation of their UK members. All $24$ communities are statistically significantly different from a random assignment, as the probability that the corresponding null model includes scientists with more diverse institutional affiliations than the actual community is one. This strongly supports the hypothesis of institutional constraint leading scientists in Business and Management to seek collaborators within institutional boundaries.

Like institutional constraint, geographic distance also plays an important role in shaping collaborations. As shown by Fig.~\ref{fig:4}d, $10$ communities exhibit statistically significantly small distances between their scientists. If the condition for significance is loosened to $P_C>0.9$, significance is even extended to $19$ communities. For a large number of communities, the probability that the average distance between all their UK members is less than randomly expected approaches one. Thus, results also provide support to the hypothesis of geographic constraint within the field of Business and Management: when scientists seek their collaborators outside their own institutions (but within the UK), they are more likely to select those who are in geographic proximity than at long distances.

In summary, the findings show that for the social scientists who submitted to the RAE 2001 in Business and Management in the UK, institutional constraint was the primary organising principle underlying their choice of scientific collaborations within the UK. Geographic constraint and status-based homophily based on the 1996 RAE rating also played a major role in shaping such collaborations, whereas research-based homophily was only marginally significant.

\section{Discussion and Conclusions}
\label{conclusion}

Prior work established that teamwork production in science is increasingly composed of collaborations that span university boundaries \cite{jones08,wuchty07}. Unlike these studies that have typically looked at institutions from multiple countries (and scientists from different disciplinary fields), our analysis has focused only on UK universities (and within a single discipline), and has suggested that scientists in Business and Management in the UK seek their collaborators within their own institutions to a greater extent than randomly expected. In this respect, our study integrates previous work on multi-university collaboration by highlighting that, when scientists' search behaviour is directed toward domestic partners within a single broad disciplinary field, it tends to remain localised within institutional boundaries. Scientists may consider collaborating with international partners \cite{jones08}; however, within their own countries and disciplinary borders, they prefer to interact with colleagues from their own institutions.

Our results also supported the role of geography in the selection of collaborators in Business and Management in the UK. Our analysis illustrated that, when collaborations span institutional boundaries, they tend to be geographically clustered. On the one hand, these findings corroborate related studies of multi-university collaborations highlighting how geographic distance can hinder group communication and decision-making \cite{cummings07}. The importance of face-to-face contacts has long been reported by the literature. Allen's \cite{allen77} rule of thumb, for example, is that collaborators should be no more than $30$ metres apart, as longer distances would negatively impact on the effectiveness of their collaboration \cite{kraut90}. On the other hand, there is an equally substantial body of literature suggesting a weakening relevance of geographic location for scientific production \cite{cairncross97,jones08}. The so-called ``death of distance'' has been mainly associated with the increasing availability of communication and computer-based technologies in research collaborations \cite{cairncross97,jones08}. Our findings complement this argument by suggesting that, when scientists choose their collaborators within their own country and discipline, they tend to favour geographic proximity. In this sense, even though the scientists included in our dataset were only partially affected by the rapid spread of information technologies in the 1990s, our results seem to suggest that technology, at least within national and disciplinary boundaries, is an imperfect substitute for geographic co-location \cite{cummings07}.

Previous research on scientific collaboration has also focused on the benefits of inter-disciplinarity, and suggested that scientists prefer collaborators from outside their own disciplinary field over those within their field \cite{laband00,whitfield08}. Since the scope of our analysis was limited only to one disciplinary field, the findings cannot provide evidence either in favour or against the tendency towards collaborations across broad disciplinary fields (e.g., physics and economics). By contrast, what they enable us to assess is the degree to which, within the boundaries of a single disciplinary field, scientists tend to collaborate across the research specialties within that field. In this respect, our results do not provide strong evidence in favour of such inter-specialty collaborations. They only partially support the hypothesis of specialty-based homophily, in that only a relatively small number of communities included scientists that were more similar in their research specialty than by chance. Since individual UK institutions inevitably tend to include only a fraction of all research specialties within Business and Management, and because scientists were found to prefer collaborations within institutional boundaries to those spanning institutions, it is not surprising to find that at least some of these collaborations occurred within the scientific boundaries of distinct specialties.

Moreover, our results provide support in favour of the signaling role of status in the choice of collaborators. In qualitative agreement with a substantial body of literature on status-based homophily \cite{chung00,lorange92,podolny94}, scientists in Business and Management were found to collaborate preferentially with others affiliated with institutions holding an RAE rating similar to the one obtained by their own institution. Similarly, recent work on multi-university research teams indicated that status is a crucial exclusivity principle underpinning scientific collaboration \cite{jones08}. These studies, for instance, reported that collaborations between top universities tend to be more common than randomly expected, especially in the social sciences. The same pattern was also found to occur between lower-tier schools, thus further intensifying the social stratification of scientific collaborations. Status therefore acts as a tangible basis for discriminating among opportunities of collaboration. Drawing on related lines of inquiry in the social sciences \cite{podolny94,podolny95}, it can be speculated that, especially when there is uncertainty about the quality of potential partners' research, the ranking of the institutions to which they belong is an attribution that scientists use to make inferences about the quality of future joint work with them. Thus, they tend to avoid partners from institutions of lower ranking than their own, and forge collaborations only with those affiliated with similarly ranked institutions. This would lead the market for collaboration to take on a ``rich-club'' structure, in which a core of scientists from top institutions form exclusive relationships with one another \cite{colizza06,opsahl08,hidalgo}.

Taken as a whole, our findings offer important insights on the underlying forces driving collaboration between scientists within a disciplinary field, and have implications for the development of mathematical models of science. Our work provides support for models going beyond a purely network point of view, and motivates the incorporation of competing non-structural factors. The importance of space on network organization is noteworthy and strongly suggests the generalization of gravity-like models \cite{frenken} in order to properly account for attractiveness over spatial distance as well as the contrary effects of the barriers between disciplines, specialties, and institutions. Similarly, the observed rich-club organization inspires the development of models where research quality across scientists and institutions is heterogeneous and constrains the way in which collaborations are forged. We believe that a precise description of these mechanisms of tie creation is crucial for predicting the emergence of complex structures such as new leading scientific communities and research teams across disciplines and specialties.

Our study is not without its limitations. First, the generalisability of the results is inevitably affected by the dataset used, with a limited geographic scope (the UK) and concerned only with a specific disciplinary field (Business and Management). Most notably, the limited scope of our dataset does not warrant generalisability of our findings to the broader domain of international and inter-disciplinary collaborations. By contrast, our analysis can only apply to collaborations involving scientists and institutions within the scientific boundaries of a single discipline and the geographic boundaries of a single country. Second, for the sake of simplicity the analysis was based only on the largest connected component of the collaboration network. Extending the analysis to other smaller connected components may well provide new insights that our analysis could not reveal. Moreover, we wish to close this section by cautioning about interpretations drawn from our method. One should indeed be careful about how our results might be influenced by the methodology, for instance our choice of community-detection algorithm. As stressed before, there exist numerous, sometimes contradictory, ways to uncover communities in networks, and we have focused here on just one particular method (i.e., optimisation of $Q(\gamma)$). More definitive conclusions about the relation between topological communities and characteristics of scientists should be drawn by comparing results obtained through different algorithms that partition the network into different communities, or even that allow scientists to belong to multiple overlapping communities. Finally, while our approach takes a purely structural viewpoint, an interesting approach would be to incorporate non-structural attributes in the definition of modules, such as more clearly hidden structural similarities between the nodes \cite{expert}.

\end{document}